\documentclass[a4paper]{article}

\usepackage{epsfig}
\usepackage{hyperref}
\usepackage{a4}
\usepackage{latexsym, amssymb} 
\usepackage{amsmath}

\newcommand{\lapprox}{%
\mathrel{%
\setbox0=\hbox{$<$}
\raise0.6ex\copy0\kern-\wd0
\lower0.65ex\hbox{$\sim$}
}}
\newcommand{\gapprox}{%
\mathrel{%
\setbox0=\hbox{$>$}
\raise0.6ex\copy0\kern-\wd0
\lower0.65ex\hbox{$\sim$}
}}

\catcode`@=11 
\def \gsim{\mathrel{\mathpalette\@versim>}}
\def \lsim{\mathrel{\mathpalette\@versim<}}
\def \@versim#1#2{\lower0.4ex\vbox{\baselineskip\z@skip\lineskip\z@skip
     \lineskiplimit\z@\ialign{$\m@th#1\hfil##\hfil$%
     \crcr#2\crcr\sim\crcr}}}
\catcode`@=12 

\makeatletter
 
 \@addtoreset{equation}{section}
\makeatother

\begin{document}

\begin{center}

{\large\bf A simple merging algorithm for jet angular correlation studies}\\[15mm]
Junya Nakamura\footnote{E-mail: junya.nakamura@uni-tuebingen.de}\\ \bigskip
{\em Institute for Theoretical Physics,
 University of T\"ubingen, \\
 Auf der Morgenstelle 14,
 72076 T\"ubingen, Germany.}
\\[20mm] 
\end{center}

\begin{abstract} 

A tree level merging algorithm which guarantees the leading order (LO) accuracy of angular correlations between jets is proposed and studied. 
The algorithm is designed so that $n$-jet events are generated exclusively according to the LO $n$-parton production cross section and each of the $n$-jet is close to each of the $n$-parton in terms of the jet measure. 
As a result, the LO accuracy of angular correlations between the $n$-jet is robust. Furthermore, as long as the $n$-jet events are exclusively subjects to a study, only the LO $n$-parton production cross section is needed and hence event generation is efficient. 
Correlations in the azimuthal angle difference between the two highest transverse momentum jets with large rapidity separations in the top quark pair production are evaluated as examples. 
The algorithm is validated by discussing numerical differences between its predictions and the predictions of a well-established merging algorithm.

\end{abstract}

\vskip 1 true cm

\newpage
\setcounter{footnote}{0}

\def\baselinestretch{1.5}
\tableofcontents


\section{Introduction}\label{sec:intro}

Angular correlations between jets produced together with heavy particles have been studied actively for a long time, because they can provide important information about the heavy particles~\cite{Plehn:2001nj, DelDuca:2001fn, Hankele:2006ma, Klamke:2007cu, Hagiwara:2009wt, Buckley:2010jv, Hagiwara:2013jp}. 
For instance it has been shown that the distribution of the azimuthal angle difference $\Delta \phi = \phi_1^{}-\phi_2^{}$ between two partons in the gluon fusion production of a Higgs boson plus the two partons is very sensitive to a charge-conjugation and parity (CP) property of the Higgs boson~\cite{Plehn:2001nj, DelDuca:2001fn, Hankele:2006ma, Klamke:2007cu, Hagiwara:2009wt}. By observing the $\Delta \phi$ distribution and comparing it with theoretical predictions, we can measure CP violation in the Higgs sector~\cite{Hankele:2006ma, Klamke:2007cu}. \\

In order to read the information of heavy particles from angular correlations between jets produced in association with them, it will be necessary to produce the accurate predictions of observables, such as $\Delta \phi$, which measure the angular correlations. Tree level merging algorithms~\cite{Catani:2001cc, Lonnblad:2001iq, Krauss:2002up, Mrenna:2003if, Lavesson:2005xu, Mangano:2006rw, Alwall:2007fs, Giele:2007di, Alwall:2008qv, Hoeche:2009rj, Hamilton:2009ne, Giele:2011cb, Lonnblad:2011xx}, which combine leading order (LO) cross sections for multi-parton in final state with the parton shower, are nowadays standard tools used for simulating processes including multi-jet in final state. Models of the parton shower base the Dokshitzer-Gribov-Lipatov-Altarelli-Parisi (DGLAP) evolution equation~\cite{Gribov:1972, Altarelli:1977, Dokshitzer:1977} and thus the parton shower guarantees the leading logarithmic (LL) accuracy for the kinematics of produced partons. Therefore the accuracy guaranteed in merging algorithms can be seen as the LO plus LL. 
The virtue of merging algorithms is that they can combine LO cross sections smoothly with the LL parton shower so that the dependence on an artificial scale at which they are combined is minimized. \\

A LO multi-parton production cross section predicts angular correlations between the produced partons at the LO accuracy, while the LL parton shower does not have ability to predict angular correlations between any partons. Considering this fact, when our objective is to predict angular correlations between constructed jets, the accuracy minimally required for the kinematics of the jets should be the LO, neither LO+LL nor LL. If the kinematics of a jet is determined or largely influenced by the LL parton shower during a merging procedure, it hardly has the LO accuracy and hence it is not appropriate to use the event containing this jet.
Merging algorithms in the literature potentially have the ambiguity in the accuracy of jets, namely it is not necessarily clear whether the kinematics of a jet constructed by clustering particles in final state after a merging procedure has the LO accuracy or not. This is because their virtue is smooth combination of LO cross sections and the LL parton shower. \\

In ref~\cite{Hagiwara:2015tva}, the azimuthal angle difference $\Delta \phi = \phi_1^{}-\phi_2^{}$ between the two highest transverse momentum $p_T^{}$ jets with a large rapidity separation in the $t\bar{t}$ pair production is studied by using the CKKW-L merging algorithm~\cite{Lonnblad:2001iq, Lavesson:2005xu, Lonnblad:2011xx} with the parton shower model~\cite{Sjostrand:2004ef, Corke:2010yf, Norrbin:2000uu} in PYTHIA8~\cite{Sjostrand:2007gs, Sjostrand:2006za}. There, it is found that the correlation between the 2-jet can be lost in a non-negligible fraction of the events when the LO cross sections for the $t\bar{t}$ plus up to 2-parton are merged with the parton shower, because a jet originating from the parton shower has a higher $p_T^{}$ than one of the two jets originating from the 2-parton of the LO cross section. 
This shows one example of the case that the kinematics of a jet is determined or largely influenced by the LL parton shower during a merging procedure. 
Although it is also found that the loss of the correlation can be avoided by merging the LO $t\bar{t}+3$-parton cross section additionally, calculating LO cross sections for higher multiplicity is time-consuming, thus we want to avoid it.\\

In this work I construct a new merging algorithm which does guarantee the LO accuracy of angular correlations between jets and hence does not have the above ambiguity. 
The Sudakov suppression is calculated in the same way as the MLM~\cite{Mangano:2006rw, Alwall:2007fs} and the $k_{\perp}^{}$-jet MLM~\cite{Alwall:2007fs, Alwall:2008qv} algorithms. The differences from these algorithms are
\begin{itemize}
\item The definition of jets used during a merging procedure is set identical to the one used during physics analyses of jets. 
\item The LO $n_{\mathrm{max}}^{}$-parton production cross section, where $n_{\mathrm{max}}^{}$ denotes the maximal number of partons produced by the LO cross section, is not allowed to produce the events which contain more than $n_{\mathrm{max}}^{}$-jet. 
\end{itemize}
The first point indicates that $n$-jet events are generated exclusively according to the LO $n$-parton production cross section and furthermore each of the $n$-jet is close to each of the $n$-parton in terms of the jet measure. As a result, angular correlations between the $n$-jet should follow those between the $n$-parton and hence the LO accuracy of them should be robust. 
In addition to this, when $n$-jet events are exclusively subjects to a study, only the LO $n$-parton production cross section is needed, thus event generation is efficient. \\

The second point makes it forbidden to produce the events which contain more than $n_{\mathrm{max}}^{}$-jet. If our objective is to predict angular correlations between two jets such as $\Delta \phi$ at the LO accuracy, not only do the two jets have angular correlations between them at the LO accuracy, but all of the additional jets also should have angular correlations with the two jets at the same accuracy, because the additional jets can affect $\Delta \phi$ kinematically. If the kinematics of the additional jets is determined by the LL parton shower, those jets do not have angular correlations with any other jets. In such a case, the accuracy of $\Delta \phi$ will be less than the LO. 
Our objective is to predict angular correlations between jets at the LO accuracy, therefore the events which contain more than $n_{\mathrm{max}}^{}$-jet should not be used for analyses. In my algorithm, this is naturally achieved by the above second point. \\

The $t\bar{t}$ production in proton-proton collisions is simulated and the azimuthal angle difference $\Delta \phi = \phi_1^{}-\phi_2^{}$ between the two highest $p_T^{}$ jets with large rapidity separations is studied by using the 2-jet events exclusively. The new algorithm is validated by comparing the $\Delta \phi$ distributions with the predictions of one of well-established algorithms, the CKKW-L algorithm~\cite{Lonnblad:2001iq, Lavesson:2005xu, Lonnblad:2011xx}. 
It is observed that the new algorithm at $n_{\mathrm{max}}^{}=2$ produces a consistent result with the CKKW-L algorithm at $n_{\mathrm{max}}^{}=3$. \\

In Section~\ref{sec:formalism}, my formalism of tree level merging algorithms and my notations are introduced. Following these, in Section~\ref{sec:new-algorithms} the new merging algorithm and an event generation procedure according to it are described in detail. In Section~\ref{sec:numerical-study}, the results of the simulation are presented. In Section~\ref{sec:conclusion}, I summarize my findings.

\section{Formalism}\label{sec:formalism}

In this section, my formalism of tree level merging algorithms and my notations are introduced. Following these, the new algorithm is described in Section~\ref{sec:new-algorithms}. This section is largely based on ref.~\cite{Hagiwara:2015tva}. \\

The Dokshitzer-Gribov-Lipatov-Altarelli-Parisi (DGLAP) evolution equation~\cite{Gribov:1972, Altarelli:1977, Dokshitzer:1977} can be written with the Sudakov form factor in the following form~\cite{Marchesini:1987cf} 
\begin{align}
t \frac{d}{dt} \frac{q(x,t)}{\Delta(t)} 
&= \int_0^{\epsilon(t)} \frac{dz}{z} \frac{\alpha_s^{}}{2\pi} \hat{P}_{qq}^{}(z)  \frac{q(x/z, t)}{\Delta(t)}, \label{dglap-with-Sudakov}
\end{align}
where $\Delta(t)$ denotes the Sudakov form factor
\begin{align}
\Delta(t)=\exp{\biggl(-\int^t_{\mu^2_{}}\frac{dt^{\prime}_{}}{t^{\prime}_{}}\int^{\epsilon(t^{\prime}_{})}_0 dz \frac{\alpha_s^{}}{2\pi}\hat{P}_{qq}^{}(z)\biggr)}. \label{Sudakov-original}
\end{align}
Here only the quark parton distribution function $q(x,t)$ (PDF) and the splitting function $\hat{P}_{qq}^{}(z)$ for $q\to qg$ without the virtual correction are introduced in order to simplify writing. The generalization is, however, simple. By integrating eq.~(\ref{dglap-with-Sudakov}) over $t_{\Lambda}^{}<t <t_X^{}$, it follows that
\begin{align}
\frac{q(x,t_X^{})}{\Delta(t_X^{})}=\frac{q(x,t_{\Lambda}^{})}{\Delta(t_{\Lambda}^{})}
+ \int^{t_X^{}}_{t_{\Lambda}^{}} \frac{dt}{t}\int^{\epsilon(t)}_0 d\hat{p}_{qq}^{}(z) \frac{q(x/z,t )}{\Delta(t)},\label{dglap-with-sudakov-integrate}
\end{align}
where a short hand notation is introduced
\begin{align}
d\hat{p}_{qq}^{}(z)= \frac{\alpha_s^{}}{2\pi} \frac{dz}{z} \hat{P}_{qq}^{}(z).\label{short-notation-1}
\end{align}
Using eq.~(\ref{dglap-with-sudakov-integrate}) iteratively and dividing it by $q(x,t_X^{})/\Delta(t_X^{})$, we can obtain the following form of the DGLAP equation~\cite{Marchesini:1987cf}
\begin{align}
1=& \frac{q(x,t_{\Lambda}^{})}{q(x,t_X^{})} \frac{\Delta(t_X^{})}{\Delta(t_{\Lambda}^{})}
+  \int^{t_X^{}}_{t_{\Lambda}^{}} \frac{dt_1^{}}{t_1^{}}\int^{\epsilon(t_1^{})}_0 d\hat{p}_{qq}^{}(z_1^{}) \frac{ q(x/z_1^{},t_{\Lambda}^{} ) }{ q(x,t_X^{}) }
\frac{\Delta(t_X^{})}{\Delta(t_{\Lambda}^{})} \nonumber \\
&+  \int^{t_X^{}}_{t_{\Lambda}^{}} \frac{dt_1^{}}{t_1^{}}\int^{\epsilon(t_1^{})}_0 d\hat{p}_{qq}^{}(z_1^{}) 
\int^{t_1^{}}_{t_{\Lambda}^{}} \frac{dt_2^{}}{t_2^{}}\int^{\epsilon(t_2^{})}_0 d\hat{p}_{qq}^{}(z_2^{}) 
\frac{ q\bigl(x/(z_1^{}z_2^{}),t_{\Lambda}^{} \bigr) }{ q(x,t_X^{}) }\frac{\Delta(t_X^{})}{\Delta(t_{\Lambda}^{})} + \cdots. \label{infinite-iteration-DGLAP-with-Sudakov-previous}
\end{align}
The first term of the right hand side (RHS) in the above equation can be seen as the probability of generating no radiation from a quark $q(x)$ in a proton during the scale evolution of the proton between $t_X^{}$ and $t_{\Lambda}^{}$ ($t_X^{}>t_{\Lambda}^{}$). The second term of the RHS represents the integrated probability of generating exclusively one radiation from the quark during the evolution, and so on. 
The left hand side (LHS) ensures the probability conservation. \\

I write the no radiation probability from a quark $q(x)$ in a proton during the scale evolution of the proton between $t_1^{}$ and $t_2^{}$ ($t_1^{}>t_2^{}$) in the following form~\cite{Marchesini:1987cf}
\begin{align}
\Pi_q^{}(t_1^{}, t_2^{}; x) = \frac{q(x,t_{2}^{})}{q(x,t_1^{})} \frac{\Delta(t_1^{})}{\Delta(t_{2}^{})}.
\end{align}
Then, eq.~(\ref{infinite-iteration-DGLAP-with-Sudakov-previous}) can be expressed as
\begin{align}
1=&
\Pi_{q}^{}( t_X^{}, t_{\Lambda}^{}; x )+
\int^{t_X^{}}_{t_{\Lambda}^{}} \frac{dt_1^{}}{t_1^{}}\int^{\epsilon(t_1^{})}_0 d\hat{p}_{qq}^{}(z_1^{})\ 
\Pi_{q}^{}( t_X^{}, t_1^{}; x)\ 
\frac{q(x/z_1^{}, t_1^{})}{q(x, t_1^{})}\ 
\Pi_q^{}( t_1^{}, t_{\Lambda}^{}; x/z_1^{} ) \nonumber \\
&
+
\int^{t_X^{}}_{t_{\Lambda}^{}} \frac{dt_1^{}}{t_1^{}}\int^{\epsilon(t_1^{})}_0 d\hat{p}_{qq}^{}(z_1^{}) 
\int^{t_1^{}}_{t_{\Lambda}^{}} \frac{dt_2^{}}{t_2^{}}\int^{\epsilon(t_2^{})}_0 d\hat{p}_{qq}^{}(z_2^{}) 
\Pi_{q}^{}( t_X^{}, t_1^{}; x)
\frac{q(x/z_1^{}, t_1^{})}{q(x, t_1^{})}
\Pi_{q}^{}( t_1^{}, t_{2}^{}; x/z_1^{} ) 
\nonumber \\
&
\hspace{0.3cm}
\times
\frac{q\bigl(x/(z_1^{}z_2^{}), t_2^{}\bigr)}{q(x/z_1^{}, t_2^{})}
\Pi_{q}^{}\bigl( t_2^{}, t_{\Lambda}^{}; x/(z_1^{}z_2^{}) \bigr) \nonumber \\
&+\cdots.\label{infinite-iteration-DGLAP-with-Sudakov}
\end{align}
It is an easy task to find the explicit form of $\Pi_q^{}(t_1^{}, t_2^{}; x)$ from the above equation~\cite{Sjostrand:1985xi}
\begin{align}
\Pi_q^{}(t_1^{}, t_2^{}; x) = \exp{\biggl( -\int^{t_1^{}}_{t_{2}^{}} \frac{dt}{t}\int^{\epsilon(t)}_0 d\hat{p}_{qq}^{}(z)\frac{q(x/z, t)}{q(x, t)} \biggr)}. \label{no-rad-prob-ISR}
\end{align}
Given the scale $t_X^{}$ and the energy fraction $x$ of an incoming quark in a proton, eq.~(\ref{infinite-iteration-DGLAP-with-Sudakov}) allows us to generate radiations from the incoming quark by evolving the proton from the scale $t_X^{}$.
This is known as backward evolution~\cite{Sjostrand:1985xi, Gottschalk:1986bk, Marchesini:1987cf}.
\\

Eq.~(\ref{infinite-iteration-DGLAP-with-Sudakov}) derived from the DGLAP equation concerns only radiation from an incoming quark in a proton i.e. initial state radiation. Here I generalize the equation to the one which can predict radiation from outgoing partons i.e. final state radiation, as well as initial state radiation.
What we should notice for this purpose is that the PDFs play a role in constraining scale evolution of the proton, or in other words constraining radiation from the quark in the proton during scale evolution of the proton. 
Hence, when radiation from outgoing partons is concerned, we replace the PDFs with a function which bases the kinematic information of the outgoing partons. The function constrains radiation from the outgoing partons, through the energy and momentum conservation for instance.\\

I let $\{p\}_{X+n}^{}$ denotes a complete specification of an event sample consisting of $X+n$ partons~\footnote{This expression is inspired by refs.~\cite{ Giele:2007di, Giele:2011cb}.}. The information of two incoming partons is implicitly included. Then, I introduce a function for the evolution of a $\{p\}_{X+n}^{}$
\begin{align}
f\bigl(z, t;  \{p\}_{X+n}^{} \bigr),\label{intro-const-function}
\end{align}
which constrains the evolution of the $\{p\}_{X+n}^{}$ at the evolution scale $t$ and the energy fraction $z$. By using the constraint function, eq.~(\ref{infinite-iteration-DGLAP-with-Sudakov}) can be generalized to 
\begin{align}
1=&
\Pi\bigl( t_X^{}, t_{\Lambda}^{}; \{p\}_X^{} \bigr)+
\int^{t_X^{}}_{t_{\Lambda}^{}} \frac{dt_1^{}}{t_1^{}}\int^1_0 d\hat{p}(z_1^{})\ 
\Pi\bigl( t_X^{}, t_1^{};  \{p\}_X^{} \bigr)\ 
f\bigl(z_1^{}, t_1^{};  \{p\}_X^{} \bigr)\ 
\Pi\bigl( t_1^{}, t_{\Lambda}^{};  \{p\}_{X+1}^{} \bigr) \nonumber \\
&
+
\int^{t_X^{}}_{t_{\Lambda}^{}} \frac{dt_1^{}}{t_1^{}}\int^1_0 d\hat{p}(z_1^{}) 
\Pi\bigl( t_X^{}, t_1^{}; \{p\}_X^{} \bigr)\ 
f\bigl(z_1^{}, t_1^{};  \{p\}_X^{} \bigr)  \nonumber \\
& \hspace{0.5cm}\times
\int^{t_1^{}}_{t_{\Lambda}^{}} \frac{dt_2^{}}{t_2^{}}\int^1_0 d\hat{p}(z_2^{}) 
\Pi\bigl( t_1^{}, t_{2}^{}; \{p\}_{X+1}^{} \bigr)
f\bigl(z_2^{}, t_2^{};  \{p\}_{X+1}^{})\ 
\Pi\bigl( t_2^{}, t_{\Lambda}^{}; \{p\}_{X+2}^{} \bigr) \nonumber \\
& + \cdots,
\label{infinite-iteration-DGLAP-with-Sudakov-general}
\end{align}
and accordingly
\begin{align}
\Pi\bigl(t_1^{}, t_2^{}; \{p\}_{X+n}^{}) = \exp{\biggl( -\int^{t_1^{}}_{t_{2}^{}} \frac{dt}{t}\int^1_0 d\hat{p}(z)f\bigl(z, t;  \{p\}_{X+n}^{}\bigr) \biggr)}, \label{no-rad-prob-general}
\end{align}
which is defined as the no radiation probability for a $\{p\}_{X+n}^{}$ as a whole, during the scale evolution of it between $t_1^{}$ and $t_2^{}$ ($t_1^{}>t_2^{}$).
In other words, this is the probability that a $\{p\}_{X+n}^{}$ remains the same during the evolution. The splitting probability has the following form
\begin{align}
d\hat{p}(z)=\frac{dz}{z}\frac{\alpha_s^{}}{2\pi}\hat{P}(z) &\mathrm{\ \ \ \ for\ initial\ state\ radiation},\nonumber \\
d\hat{p}(z)=dz\frac{\alpha_s^{}}{2\pi}\hat{P}(z) &\mathrm{\ \ \ \   for\ final\ state\ radiation}, \label{short-hand-splitting-function}
\end{align}
where the appropriate splitting function(s) should be used for $\hat{P}(z)$ according to a branching process $\{p\}_{X+n}^{} \to \{p\}_{X+(n+1)}^{}$. For initial state radiation, the constraint function $f(z, t;  \{p\}_{X+n}^{})$ always includes the PDFs. The soft gluon singularity at $z=1$ in the splitting functions will be avoided by introducing $\theta$ functions in the constraint function. \\

Let us consider a hard process which produces $X$ and express its leading order (LO) cross section as $\sigma(X)$. The DGLAP evolution of the hard process can be expressed by multiplying $\sigma(X)$ by eq.~(\ref{infinite-iteration-DGLAP-with-Sudakov-general}),
\begin{align}
\sigma(X)
&=\sigma(X)\ \Pi\bigl( t_X^{}, t_{\Lambda}^{}; \{p\}_X^{} \bigr) \nonumber \\
&+\sigma(X) \int^{t_X^{}}_{t_{\Lambda}^{}} \frac{dt_1^{}}{t_1^{}}\int^1_0 d\hat{p}(z_1^{})\ 
\Pi\bigl( t_X^{}, t_1^{};  \{p\}_X^{} \bigr)\ 
f\bigl(z_1^{}, t_1^{};  \{p\}_X^{} \bigr)\ 
\Pi\bigl( t_1^{}, t_{\Lambda}^{};  \{p\}_{X+1}^{} \bigr) \nonumber \\
&+\sigma(X) \int^{t_X^{}}_{t_{\Lambda}^{}} \frac{dt_1^{}}{t_1^{}}\int^1_0 d\hat{p}(z_1^{}) 
\Pi\bigl( t_X^{}, t_1^{}; \{p\}_X^{} \bigr)\ 
f\bigl(z_1^{}, t_1^{};  \{p\}_X^{} \bigr)  \nonumber \\
& \hspace{0.5cm}\times
\int^{t_1^{}}_{t_{\Lambda}^{}} \frac{dt_2^{}}{t_2^{}}\int^1_0 d\hat{p}(z_2^{}) 
\Pi\bigl( t_1^{}, t_{2}^{}; \{p\}_{X+1}^{} \bigr)
f\bigl(z_2^{}, t_2^{};  \{p\}_{X+1}^{})\ 
\Pi\bigl( t_2^{}, t_{\Lambda}^{}; \{p\}_{X+2}^{} \bigr) \nonumber \\
& + \cdots.\label{X-DGLAP-evolution}
\end{align}
Tree level merging algorithms are designed as to improve the above DGLAP evolution by replacing the terms constructed by the leading order cross section times the universal radiation probability with the exact LO cross sections for multi-parton in final state~\cite{Catani:2001cc}. In my notations, it proceeds as
\begin{align}
\sigma(X)
&=\sigma(X)\ \Pi\bigl( t_X^{}, t_{\Lambda}^{}; \{p\}_X^{} \bigr) \nonumber \\
&+\sigma\bigl(X+1; \{p\}_{X+1}^{} > t_{\Lambda}^{} \bigr) \ 
\Pi\bigl( t_X^{}, t_1^{}; \{p\}_X^{} \bigr)\ 
f^{\prime}_{}\bigl(z_1^{}, t_1^{};  \{p\}_X^{} \bigr)\ 
\Pi\bigl( t_1^{}, t_{\Lambda}^{}; \{p\}_{X+1}^{} \bigr) \nonumber \\
&+\sigma\bigl(X+2; \{p\}_{X+2}^{} > t_{\Lambda}^{} \bigr) \
\Pi\bigl( t_X^{}, t_1^{}; \{p\}_X^{} \bigr)\ 
f^{\prime}_{}\bigl(z_1^{}, t_1^{};  \{p\}_X^{} \bigr)\
\Pi\bigl( t_1^{}, t_{2}^{}; \{p\}_{X+1}^{} \bigr) \nonumber \\
&
\hspace{0.5cm}
\times f^{\prime}_{}\bigl(z_2^{}, t_2^{};  \{p\}_{X+1}^{})\ 
\Pi\bigl( t_2^{}, t_{\Lambda}^{}; \{p\}_{X+2}^{} \bigr)\ \nonumber \\
& + \cdots,\label{improved-DGLAP-1}
\end{align}
The soft and collinear divergences in LO cross sections are regularized in the following way. By using the definition of the evolution variable $t$, calculate the minimum value $t_{\mathrm{min}}^{}$ from a $\{p\}_{X+n}^{}$ and require $t_{\mathrm{min}}^{}>t_{\Lambda}^{}$. This is expressed as $\{p\}_{X+n}^{}>t_{\Lambda}^{}$ in the above equation~\footnote{Here it is assumed that the hard process cross section $\sigma(X)$ is finite everywhere in its phase space.}. Notice that the constraint functions in eq.~(\ref{improved-DGLAP-1}) are different from those in eq.~(\ref{X-DGLAP-evolution}), since some part of the constraint is already included in the exact LO cross section. I call $f^{\prime}_{}(z_{n+1}^{}, t_{n+1}^{};  \{p\}_{X+n}^{})$ a weight function hereafter, since it is used to re-weight an event sample during a merging procedure.
The cut off scale $t_{\Lambda}^{}$ is called merging scale and separates phase space into two regions. Partons in the phase space above the merging scale are produced following exact LO cross sections and those in the phase space below the merging scale are produced following the DGLAP equation. The further evolution of each term in the above equation will be performed between the scale $t_{\Lambda}^{}$ and a lower cutoff scale for complete event generation. In my notations, the term consisting of the cross section $\sigma(X+n)$ in the RHS of the above equation is called the $n$ th term. \\

Below tree level merging algorithms which have been proposed and studied in the literature are briefly reviewed. This information can be useful when I describe my algorithm in Section~\ref{sec:new-algorithms}. 
There are fundamentally two different types of merging algorithms~\footnote{In this work I concentrate only on discussing a type of merging algorithms which introduces the merging scale. However, it would be also interesting to study jet angular correlations with other types of algorithms such as \cite{Giele:2011cb}.}. One of the two is constructed to generate events according to the improved DGLAP equation strictly, eq.~(\ref{improved-DGLAP-1}) in my notations, by explicitly calculating the Sudakov form factors. This procedure is necessary in order to minimize the dependence on the merging scale and smoothly combine LO cross sections with the parton shower.
Algorithms of this type include the CKKW with the vetoed parton shower~\cite{Catani:2001cc, Krauss:2002up}, the CKKW with the truncated shower~\cite{Hoeche:2009rj, Hamilton:2009ne} and the CKKW-L~\cite{Lonnblad:2001iq, Lavesson:2005xu, Lonnblad:2011xx} algorithms.
The other of the two types does not calculate the Sudakov form factors explicitly and hence event generation follow the improved DGLAP equation approximately. 
Algorithms of this type include the MLM~\cite{Mangano:2006rw, Alwall:2007fs}, the $k_{\perp}^{}$-jet MLM~\cite{Alwall:2007fs, Alwall:2008qv} and the shower $k_{\perp}^{}$~\cite{Alwall:2008qv} algorithms. In the MLM and the $k_{\perp}^{}$-jet MLM algorithms, partons in final state obtained after the complete shower evolution are clustered according to the cone jet algorithm and the exclusive $k_{\perp}^{}$-jet algorithm, respectively. Then, if the number of the constructed jets is not equal to the number of partons produced by a LO cross section or if the constructed jets are not close to the partons in terms of the jet measure, the event is vetoed. In the shower $k_{\perp}^{}$ algorithm, if the scale of the first emission is above the merging scale, then the event is vetoed.

\section{Algorithm}\label{sec:new-algorithms}

In this section, the ideas of the new merging algorithm and an event generation procedure according to the algorithm are described in detail.\\

The new algorithm is constructed with a goal of eliminating the ambiguity in the accuracy of jets which potentially exists in merging algorithms as discussed in Section~\ref{sec:intro}, and hence of guaranteeing the leading order (LO) accuracy of angular correlations between jets. The following idea is strictly implemented in the algorithm:\\

{\it The $n$-jet events are generated exclusively according to the LO $n$-parton production cross section and furthermore each of the $n$-jet is close to each of the $n$-parton in terms of the jet measure.}\\

For an algorithm which implements the above idea, which I call the model A algorithm~\footnote{This is the remnant of the fact that there existed the model B too, which turned out not to work properly and thus is not described in this paper.}, eq.~(\ref{improved-DGLAP-1}) may be written in the following form:
\begin{align}
&\sigma(X)_{\mathrm{Model A}}^{} \nonumber\\
&=\sigma(X)\ \Pi\bigl( t_X^{}, \{p\}_{X}^{} \to \{p\}_{X+0\mathrm{jet}}^{} ; \{p\}_X^{} \bigr) \nonumber \\
&+\sigma\Bigl(X+1; \{p\}_{X+1}^{} > Q_{\mathrm{cut}}^{\mathrm{ME}} \Bigr) \ 
\Pi\bigl( t_{X+1}^{}, \{p\}_{X+1}^{} \to \{p\}_{X+1\mathrm{jet}}^{} ; \{p\}_{X+1}^{} \bigr) \nonumber \\
&+\sigma\Bigl(X+2; \{p\}_{X+2}^{} > Q_{\mathrm{cut}}^{\mathrm{ME}} \Bigr) \ 
\Pi\bigl( t_{X+2}^{}, \{p\}_{X+2}^{} \to \{p\}_{X+2\mathrm{jets}}^{} ; \{p\}_{X+2}^{} \bigr) \nonumber \\
& + \cdots \nonumber \\
& + \sigma\Bigl(X+n; \{p\}_{X+n}^{} > Q_{\mathrm{cut}}^{\mathrm{ME}} \Bigr) \ 
\Pi\bigl( t_{X+n}^{}, \{p\}_{X+n}^{} \to \{p\}_{X+n\mathrm{jets}}^{} ; \{p\}_{X+n}^{} \bigr) \nonumber \\
& + \cdots \nonumber \\
& + \sigma\Bigl(X+n_{\mathrm{max}}^{}; \{p\}_{X+n_{\mathrm{max}}^{}}^{} > Q_{\mathrm{cut}}^{\mathrm{ME}} \Bigr) \ 
\Pi\bigl( t_{X+n_{\mathrm{max}}^{}}^{}, \{p\}_{X+n_{\mathrm{max}}^{}}^{} \to \{p\}_{X+n_{\mathrm{max}}^{} \mathrm{jets}}^{} ; \{p\}_{X+n_{\mathrm{max}}^{}}^{} \bigr) .\label{improved-DGLAP-2}
\end{align}
Here the weight functions are omitted in order to simplify writing. In the numerical studies in Section~\ref{sec:numerical-study}, they are taken into account. 
The expression $\{p\}_{X+n}^{} > Q_{\mathrm{cut}}^{\mathrm{ME}}$ indicates that the soft and collinear divergences in LO cross sections are regularized by the definition of $Q_{\mathrm{cut}}^{\mathrm{ME}}$. 
The definition of $Q_{\mathrm{cut}}^{\mathrm{ME}}$ must respect the definition of jets used in analyses, namely a jet clustering algorithm and parameters that the algorithm contains. 
Throughout of this work, the anti-$k_T^{}$ algorithm~\cite{Cacciari:2008gp} is used as a jet clustering algorithm, thus I concentrate on discussing the case of using the anti-$k_T^{}$ algorithm. 
The anti-$k_T^{}$ algorithm basically contains two parameters which we can choose their values freely, namely the radius parameter $R^{\mathrm{jet}}_{}$ and the lower transverse momentum cutoff on jets $p_{T \mathrm{cut}}^{\mathrm{jet}}$. Hence $Q_{\mathrm{cut}}^{\mathrm{ME}}$ should be defined as 
\begin{subequations}\label{ME-cutoff}
\begin{align}
\Delta R_{ij}^{} =\sqrt{(y_i^{}-y_j^{})^2_{}+(\phi_i^{}-\phi_j^{})^2_{}} > R_{\mathrm{cut}}^{\mathrm{ME}},\label{R-para}\\
p_{T i}^{} > p_{T \mathrm{cut}}^{\mathrm{ME}},
\end{align}
\end{subequations}
where $p_{T i}^{}$, $y_i^{}$ and $\phi_i^{}$ are the transverse momentum with respect to the beam, rapidity and azimuthal angle of outgoing parton $i$. Imposing cutoffs on the rapidity of partons in not needed.  
In order to avoid missed phase space, the above two parameters must satisfy
\begin{align}
p_{T \mathrm{cut}}^{\mathrm{jet}} \ge p_{T \mathrm{cut}}^{\mathrm{ME}},\ \ 
R^{\mathrm{jet}}_{} \ge R_{\mathrm{cut}}^{\mathrm{ME}}.\label{mergin-scale-rel}
\end{align}
\\

Let us consider generating an $n$-jet event according to the $n$ th term in eq.~(\ref{improved-DGLAP-2}),
\begin{align}
\sigma\Bigl(X+n; \{p\}_{X+n}^{} > Q_{\mathrm{cut}}^{\mathrm{ME}} \Bigr) \ 
\Pi\bigl( t_{X+n}^{}, \{p\}_{X+n}^{} \to \{p\}_{X+n\mathrm{jets}}^{} ; \{p\}_{X+n}^{} \bigr).
\end{align}
A parton shower program is executed on an event sample $\{p\}_{X+n}^{}$ generated by the LO cross section $\sigma(X+n; \{p\}_{X+n}^{} > Q_{\mathrm{cut}}^{\mathrm{ME}})$, by setting the shower starting scale to $t_{X+n}^{}$. 
Once the shower evolution is performed until the shower cutoff scale, all partons in the final state within a rapidity range $|y|< y_{\mathrm{cut}}^{\mathrm{clus}}$ are clustered to construct jets according to the anti-$k_T^{}$ algorithm with the radius parameter $R^{\mathrm{jet}}_{}$ and the lower $p_T^{}$ cutoff $p_{T \mathrm{cut}}^{\mathrm{jet}}$. If the number of the constructed jets $n_{\mathrm{jet}}^{}$ is not identical to $n$, the event sample is vetoed. If the event sample survives, the distance parameters $\Delta R$ defined in eq.~(\ref{R-para}) between the $n$-jet and the $n$-parton are calculated, and then it is checked whether the following is satisfied between each of the $n$-jet and each of the $n$-parton, or not
\begin{align}
\Delta R_{\mathrm{jet,\ parton}}^{} < C_{\mathrm{match}}^{}\times R^{\mathrm{jet}}_{}. \label{C-match-parameter}
\end{align}
If a jet satisfies the above relation with a parton, it means that the jet is close to the parton in terms of the measure of the anti-$k_T^{}$ algorithm. The jet is called matched with the parton. 
If a matching between a jet and a parton is confirmed for all of the $n$-jet and all of the $n$-parton, then the event sample is accepted. Once the event sample is accepted, all partons in the final state are again clustered to construct jets which are defined in the same way as above, but this time only those within a rapidity range $|y|< y_{\mathrm{cut}}^{\mathrm{detect}}$ are clustered. A value for $y_{\mathrm{cut}}^{\mathrm{detect}}$ should reflect actual detectors at experiments and can be different from $y_{\mathrm{cut}}^{\mathrm{clus}}$ which is used during the merging procedure. The following relation should be satisfied between the two
\begin{align}
y_{\mathrm{cut}}^{\mathrm{detect}} \le y_{\mathrm{cut}}^{\mathrm{clus}}.
\end{align}
If values for these two cuts are different, the jets constructed at this stage cannot be identical to those constructed during the merging procedure and thus are not necessarily matched with the $n$-parton in some events. However, if $y_{\mathrm{cut}}^{\mathrm{detect}}$ is large enough so that there is only a small fraction of the events $\{p\}_{X+n}^{}$ above $y_{\mathrm{cut}}^{\mathrm{detect}}$, then the effect is expected to be small. In the numerical studies, I use $y_{\mathrm{cut}}^{\mathrm{detect}}=5$ and $y_{\mathrm{cut}}^{\mathrm{clus}}=7$. The constructed jets will be used for physics analyses. \\

The accepted events have $n$-jet exclusively and each of the $n$-jet is close to each of the $n$-parton produced by the LO cross section $\sigma(X+n)$ in terms of the jet measure. Angular correlations between the $n$-jet should follow those between the $n$-parton. Since angular correlations between the $n$-parton is at the LO accuracy, those between the $n$-jet should also be at the LO accuracy. \\

As is clear, the algorithm calculates the effects of the Sudakov form factors in the same way as the MLM~\cite{Mangano:2006rw, Alwall:2007fs} and the $k_{\perp}^{}$-jet MLM~\cite{Alwall:2007fs, Alwall:2008qv} algorithms~\footnote{See the discussion at the end of Section~\ref{sec:formalism}.}. The differences from these algorithms are
\begin{itemize}
\item The definition of jets used during a merging procedure is set identical to the one used during physics analyses of jets. 
\item The LO $n_{\mathrm{max}}^{}$-parton production cross section is not allowed to produce the events which contain more than $n_{\mathrm{max}}^{}$-jet. 
\end{itemize}
As to the first point, the MLM and the $k_{\perp}^{}$-jet MLM algorithms use the cone algorithm and the exclusive $k_{\perp}^{}$-jet algorithm, respectively, as a clustering algorithm, and parameters that the clustering algorithm contains are chosen independently of the definition of jets used in physics analyses. In other words, the definition of jets used in physics analyses is nothing to do with a merging procedure in these merging algorithms. In my algorithm, the first point is essential in order to guarantee the LO accuracy of angular correlations between jets. \\

The maximal number $n_{\mathrm{max}}^{}$ of partons produced by the LO cross section is limited. If we follow the treatment of the LO $n_{\mathrm{max}}^{}$-parton production cross section in the MLM and the $k_{\perp}^{}$-jet MLM algorithms, the cross section is allowed to produce the events which contain $n_{\mathrm{max}}^{}$ or more than $n_{\mathrm{max}}^{}$-jet and each of the hardest $n_{\mathrm{max}}^{}$-jet is required to be matched with each of the $n_{\mathrm{max}}^{}$-parton. In this approach, the kinematics of the additional jets i.e. jets softer than the $n_{\mathrm{max}}^{}$-jet in terms of the jet measure is determined by the leading logarithmic (LL) parton shower and hence they do not have angular correlations with any other jets. 
Since our objective is to predict angular correlations between jets at the LO accuracy, not only do the jets used for an observable such as $\Delta \phi=\phi_1^{}-\phi_2^{}$ have angular correlations between them at the LO accuracy, but all of the jets including jets not used for the observable should have angular correlations between them at the same accuracy, because even the jets not used for the observable can affect the observable kinematically. If some of the jets not used for the observable are generated by the LL parton shower, then the accuracy of the observable will be less than the LO. Considering this argument, the treatment of the LO $n_{\mathrm{max}}^{}$-parton production cross section in the MLM and the $k_{\perp}^{}$-jet MLM algorithms is not appropriate to our objective. In my algorithm, the second point above guarantees the LO accuracy of angular correlations between jets in all events. \\

The parameter $C_{\mathrm{match}}^{}$ in eq.~(\ref{C-match-parameter}) is an important parameter, since the Sudakov suppression can depend on it. 
The implementations of the MLM algorithm in Alpgen~\cite{Mangano:2006rw, Alwall:2007fs} and in MadGraph5\verb|_|aMC@NLO~\cite{Alwall:matching} use $C_{\mathrm{match}}^{}=1.5$.
However there is nothing that uniquely determines $C_{\mathrm{match}}^{}$, thus which should be considered as a tuning parameter and can also depend on a jet clustering algorithm. In my implementation of the new algorithm with the anti-$k_T^{}$ algorithm, $C_{\mathrm{match}}^{}=2.0$ is used. \\

Below I describe a procedure of event generation for completeness. The numerical studies in Section~\ref{sec:numerical-study} are performed according to this.
\begin{enumerate}
\item Generate the event samples for the $X+0, 1, \dots, n_{\mathrm{max}}^{}$-parton production processes at proton-proton (pp) collisions according to the LO cross sections, i.e. $\{p\}_{X}^{}, \{p\}_{X+1}^{}, \cdots, \{p\}_{X+n_{\mathrm{max}}^{}}^{}$. Here the $t\bar{t}$ production is simulated, thus $X=t\bar{t}$.
MadGraph5\verb|_|aMC@NLO\cite{Alwall:2014hca} version 5.2.2.1 is used for this purpose. The soft and collinear singularity is regularized by imposing eq.~(\ref{ME-cutoff}). I use
\begin{align}
R_{\mathrm{cut}}^{\mathrm{ME}}=0.4,\ \ p_{T \mathrm{cut}}^{\mathrm{ME}}=20\ \mathrm{GeV}.\label{ME-cut-value}
\end{align}
A fixed value $t_{\Lambda}^{}$ is used for the scales in the strong couplings and in the parton distribution functions (PDFs). The PDF set CTEQ6L1~\cite{Pumplin:2002vw} is used. Note that, when the $X+n$-jet events are exclusively subjects to a study, only the events $\{p\}_{X+n}^{}$ are needed to be generated.

\item Select an event sample for the $X+n$ partons process, i.e. $\{p\}_{X+n}^{}$,  with the probability proportional to its integrated LO cross section obtained in step 1,
\begin{align}
P_n=\frac{\sigma(pp\to X+n)}{\sum_{i=0}^{n_{\mathrm{max}}^{}}\sigma(pp\to X+i)}.
\end{align}

\item Construct a PYTHIA8 parton shower history of the $\{p\}_{X+n}^{}$. The history consists of intermediate events $\{p\}_{X+(n-1)}^{}, \{p\}_{X+(n-2)}^{}, \cdots, \{p\}_{X+i}^{}, \cdots, \{p\}_{X+1}^{}, \{p\}_{X}^{}$ with the clustering scales $t_n^{} < t_{n-1}^{} < \cdots < t_{i+1}^{} < \cdots < t_2^{}< t_1^{}$.

\item Calculate the weight function for the $\{p\}_{X+n}^{}$. Let us define the energy fractions and the parton types of the incoming partons in the $\{p\}_{X+i}^{}$ by $x_1^{(i)}$, $x_2^{(i)}$ and $f_1^{(i)}$, $f_2^{(i)}$, respectively. The weight function for the $\{p\}_{X+i}^{}$ is given by~\cite{Lonnblad:2011xx}
\begin{align}
f^{\prime}_{}\bigl(z_{i+1}^{}, t_{i+1}^{};  \{p\}_{X+i}^{} \bigr)
=
\frac{ \alpha_s^{}( t_{i+1}^{} ) }{ \alpha_s^{}( t_{\Lambda}^{} ) } 
\frac{ f_1^{(i)}( x_1^{(i)}, t_i^{} ) }{ f_1^{(i)}( x_1^{(i)}, t_{i+1}^{} ) }
\frac{ f_2^{(i)}( x_2^{(i)}, t_i^{} ) }{ f_2^{(i)}( x_2^{(i)}, t_{i+1}^{} ) }, \label{general-const-function}
\end{align}
from which the total weight function for the $\{p\}_{X+n}^{}$ is 
\begin{align}
\prod_{i=0}^{n}f^{\prime}_{}\bigl(z_{i+1}^{}, t_{i+1}^{};  \{p\}_{X+i}^{} \bigr),\label{total-const-function}
\end{align}
where $t_0^{}=t_X^{}$ and $t_{n+1}^{}=t_{\Lambda}^{}$. The PYTHIA8 shower model uses the different strong couplings for initial state radiation and final state radiation. Therefore, when a radiation is classified as the initial state radiation (the final state radiation) with a scale $t_{i+1}^{}$ by the shower history construction,  $\alpha_s^{}(m_z^{})=1.37$ $(1.383)$ is used for the factor in eq.~(\ref{general-const-function}). 
The scale $t_X^{}$ should be determined from the intermediate event $\{p\}_{X}^{}$. I define it by
\begin{align}
t_X^{}=E_T^{}(t) \times E_T^{}(\bar{t}),\label{tX-define}
\end{align}
where $E_T^2=m^2_{}+p_T^2$. This scale $t_X^{}$ is also used as the renormalization scales of the strong couplings $\alpha_s^2$ for the hard process, namely
\begin{align}
\frac{\alpha_s^2(t_X^{}) }{ \alpha_s^2(t_{\Lambda}^{}) }
\end{align}
is added as a multiplicative factor in the weight function. I use $\alpha_s^{}(m_z^{})=0.13$ in the above factor. Once the weight function is calculated, the $\{p\}_{X+n}^{}$ is re-weighted with the function. However, since the weight function is not bounded above by unity, the upper bound of the weight function must be found at first by calculating the weight function for a large number of events $\{p\}_{X+n}^{}$. The integrated LO cross section obtained in step 1 has to be multiplied by the obtained upper bound of the weight function.

\item Calculate the effects of the Sudakov form factors by using the method described above. I use the parton shower model~\cite{Sjostrand:2004ef, Corke:2010yf, Norrbin:2000uu} in PYTHIA8~\cite{Sjostrand:2007gs, Sjostrand:2006za} version 8186. The default tune of the version 8186, tune 4C~\cite{Corke:2010yf}, is used. 
The parameters in the anti-$k_T^{}$ algorithm and the cutoff values on the rapidity of partons are set to
\begin{align}
R^{\mathrm{jet}}_{}=0.4,\ \ p_{T \mathrm{cut}}^{\mathrm{jet}}=30\ \mathrm{GeV},\ \  y_{\mathrm{cut}}^{\mathrm{clus}}=7,\ \ y_{\mathrm{cut}}^{\mathrm{detect}}=5.
\end{align}
I use Fastjet~\cite{Cacciari:2011ma} version $3.1.0$ for executing the anti-$k_T^{}$ algorithm. 
The parton shower starting scale, which is the maximal shower evolution scale and is denoted by $t_{X+n}^{}$ in eq.~(\ref{improved-DGLAP-2}), is set to
\begin{align}
t_{X+n}^{} = t_X^{}/4 + t_1^{}/16 + t_2^{}/16 + \cdots + t_n^{}/16.\label{starting-scale}
\end{align}

\item Repeat the above procedures from step 2 to step 5 until a large number of the accepted event samples are obtained. 

\end{enumerate}

\section{Numerical studies}\label{sec:numerical-study}

In this section, the model A algorithm is validated by discussing numerical differences between its predictions and the predictions of another merging algorithm. As another merging algorithm, I choose the CKKW-L merging algorithm~\cite{Lonnblad:2001iq, Lavesson:2005xu, Lonnblad:2011xx}, which is one of well-established algorithms~\footnote{In my implementation of the CKKW-L algorithm, the method of phase space separation is different from the original one. This is the one called the CKKW-L+ algorithm in ref.~\cite{Hagiwara:2015tva}. However numerical differences are found small.}. 
The top quark pair production process at the 14 TeV LHC is simulated, and then angular correlations between the two highest transverse momentum jets are studied by using the $t\bar{t}+2$-jet events exclusively. 
The merging scale in the CKKW-L algorithm is set as eq.~(\ref{ME-cut-value}). \\

As an observable measuring angular correlations between the two hardest jets i.e. the two highest transverse momentum jets, the azimuthal angle difference is chosen
\begin{align}
\Delta \phi = \phi_1^{}-\phi_2^{}.
\end{align}
The events which contain 2-jet are exclusively picked up at first and then if the 2-jet pass the following rapidity cut, $\Delta \phi$ is evaluated by using the 2-jet,
\begin{align}
y_1^{} \times y_2^{} < 0,\ \ \ \ |y_1^{}-y_2^{}|>3.5\ \ \mathrm{or}\ \ 2.5, \label{vbf}
\end{align}
which are often called the vector boson fusion (VBF) cuts. One of the 2-jet which has a positive rapidity is chosen for $\phi_1^{}$ and the other jet which has a negative rapidity is chosen for $\phi_2^{}$.
The studies based on the leading order (LO) $t\bar{t}+2$-parton production cross section~\cite{Buckley:2010jv, Hagiwara:2013jp} have shown that there are strong correlations between the two partons with a large rapidity separation, when the $t\bar{t}$ is near or far away from the threshold of its production~\cite{Hagiwara:2013jp}. Here I do not impose any cuts on the $t\bar{t}$, however. 
The rapidity cut on jets is set as $|y| < 4.5$. The $t\bar{t}$ is assumed stable, since the main purpose of this study is to investigate a way of accurately modeling the kinematic activity of jets induced by the hard process. Hence the hadronization after the shower evolution and the multiple interaction in PYTHIA8 are also turned off. \\

\begin{figure}[t]
\centering
\includegraphics[scale=0.54]{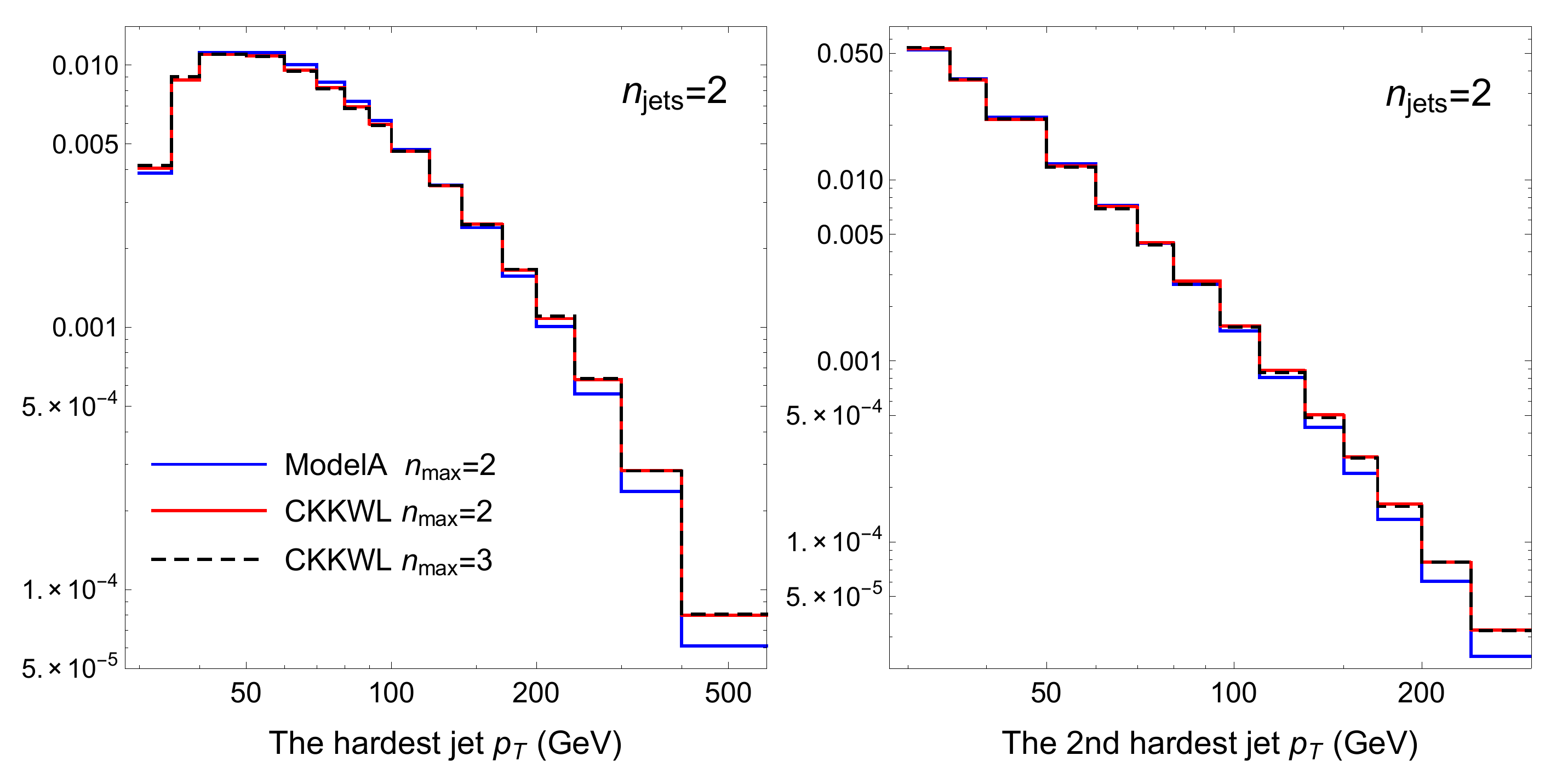}
\caption{\small
The normalized differential cross section as a function of the $p_T^{}$ of the highest $p_T^{}$ jet (left panel) and that of the second highest $p_T^{}$ jet (right panel) in the 2-jet events.  
The correspondence between the curves and the merging algorithms is labeled inside the left panel.
}
\label{figure:jet-dist}
\end{figure}

At first, I show the transverse momentum $p_T^{}$ distributions of jets. 
In the left and the right panels of Figure~\ref{figure:jet-dist}, 
the normalized differential cross section as a function of the transverse momentum $p_T^{}$ of the highest $p_T^{}$ jet and that of the second highest $p_T^{}$ jet in the 2-jet events are shown, respectively. 
The blue solid curve represents the distribution of the model A algorithm. 
The red solid curve and the black dashed curve represent the distributions of the CKKW-L algorithm, the maximal number $n_{\mathrm{max}}^{}$ of partons produced by the LO cross section is $n_{\mathrm{max}}^{}=2$ for the former and $n_{\mathrm{max}}^{}=3$ for the latter. \\

The important difference of the two algorithms is that only the $2$-parton production cross section contributes to the 2-jet events in the model A algorithm, while the $0$, $1$ and $2$-parton production cross sections contribute to the 2-jet events in the CKKW-L algorithm at $n_{\mathrm{max}}^{}=2$ and the $0$, $1$, $2$ and $3$-parton production cross sections contribute to the 2-jet events in the CKKW-L algorithm at $n_{\mathrm{max}}^{}=3$~\footnote{The contribution of the $0$ and $1$-parton cross sections to the the 2-jet events is less than $5\%$, since the merging scale is set as eq.~(\ref{ME-cut-value}).}. \\

The panels show that the model A algorithm produces softer distributions than the CKKW-L algorithm, and that the CKKW-L at $n_{\mathrm{max}}^{}=2$ and the CKKW-L at $n_{\mathrm{max}}^{}=3$ produce consistent distributions. 
Since the two algorithms calculate the Sudakov form factors in the very different ways, the differences in the $p_T^{}$ distributions are not unexpected. The model A algorithm contains a tuning parameter $C_{\mathrm{match}}^{}$ in eq.~(\ref{C-match-parameter}). In addition to this, the parton shower starting scale, which is defined as in eq.~(\ref{starting-scale}), is not uniquely determined. Therefore, we may tune the algorithm by using these parameters so that the $p_T^{}$ distributions become consistent with the experimental data, before predicting $\Delta \phi$ distributions. This issue is discussed again in Section~\ref{sec:conclusion}. \\

\begin{figure}[t]
\centering
\includegraphics[scale=0.52]{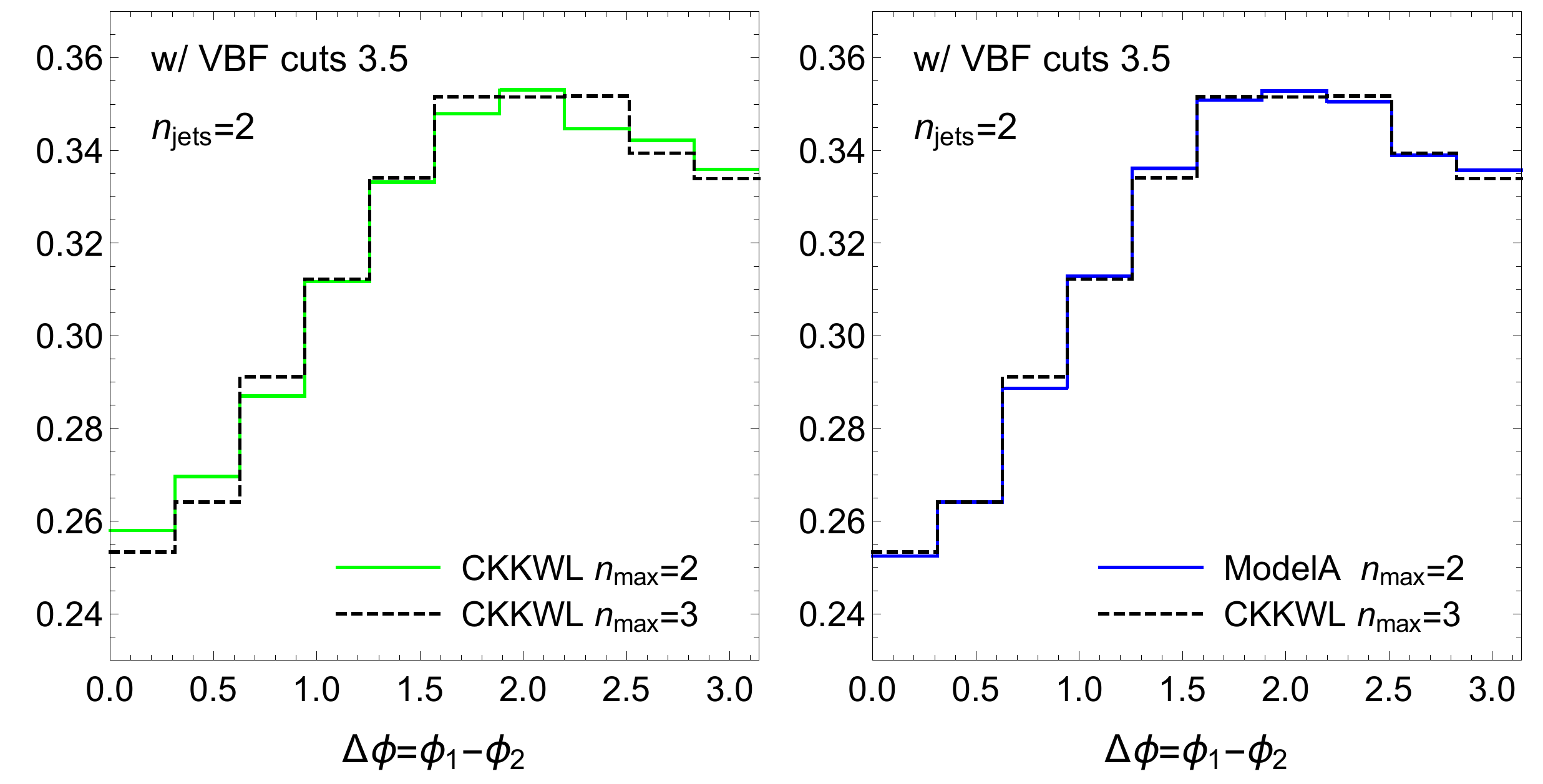}
\includegraphics[scale=0.52]{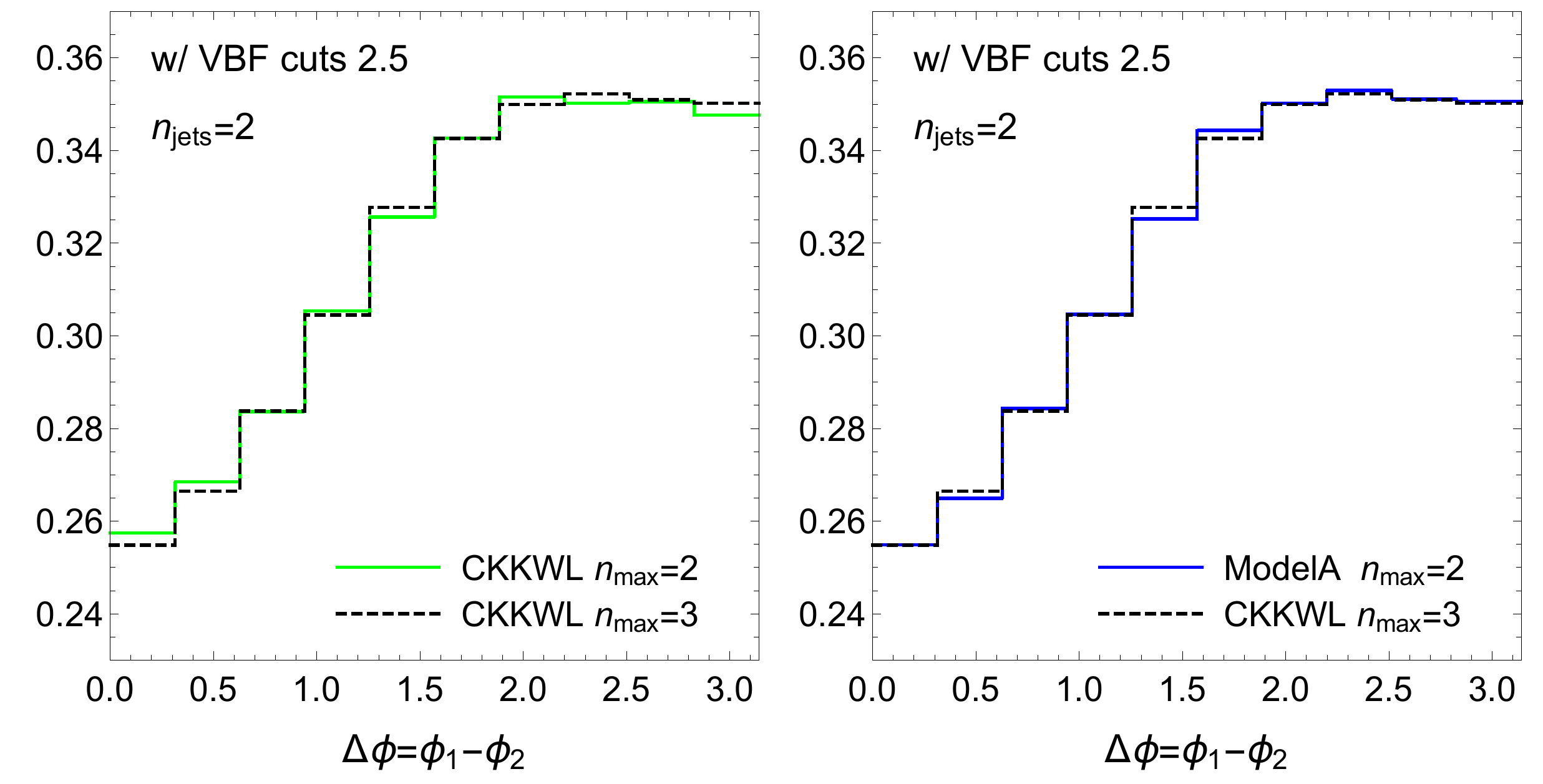}
\caption{\small 
The normalized differential cross section as a function of $\Delta \phi$ with the rapidity separation $|y_1^{}-y_2^{}|>3.5$ (upper two panels) and $|y_1^{}-y_2^{}|>2.5$ (lower two panels) in the 2-jet events. The correspondence between the curves and the merging algorithms are shown inside the panels.}
\label{figure:dphi-dist}
\end{figure}

In Figure~\ref{figure:dphi-dist} the $\Delta \phi$ distributions are shown. The normalized differential cross section as a function of $\Delta \phi$ with the rapidity separation $|y_1^{}-y_2^{}|>3.5$ and that as a function of $\Delta \phi$ with the rapidity separation $|y_1^{}-y_2^{}|>2.5$ are shown in the upper two panels and the lower two panels, respectively. The correspondence between the curves and the merging algorithms are shown inside the panels. \\

In the left panels of Figure~\ref{figure:dphi-dist}, the CKKW-L at $n_{\mathrm{max}}^{}=2$ and the CKKW-L at $n_{\mathrm{max}}^{}=3$ are compared. The difference between the two predictions, particularly in the upper left panel, can be explained from a loss of the correlation due to the contribution from the leading logarithmic (LL) parton shower in the CKKW-L at $n_{\mathrm{max}}^{}=2$~\cite{Hagiwara:2015tva}. \\

In the right panels of Figure~\ref{figure:dphi-dist}, the model A algorithm and the CKKW-L at $n_{\mathrm{max}}^{}=3$ are compared. 
It is shown that the model A algorithm produces a consistent prediction with the CKKW-L at $n_{\mathrm{max}}^{}=3$, despite that only the LO $t\bar{t}+2$-parton cross section contributes to the events in the model A algorithm. 
Since the LO accuracy of $\Delta \phi$ is guaranteed in the model A algorithm, this observation confirms the finding in ref.~\cite{Hagiwara:2015tva} that the LO accuracy of $\Delta \phi$ in the 2-jet events cannot be fully retained due to the contribution from the LL parton shower in the CKKW-L at $n_{\mathrm{max}}^{}=2$ and it can be retained by merging the LO $t\bar{t}+3$-parton cross section additionally i.e. the CKKW-L at $n_{\mathrm{max}}^{}=3$. 
It should be worth emphasizing that, in the model A algorithm, the LO $t\bar{t}+3$-parton cross section never contributes the $\Delta \phi$ distributions in Figure~\ref{figure:dphi-dist}, since it contributes exclusively to the 3-jet events.

\section{Summary and discussion}\label{sec:conclusion}

In this paper, a new tree level merging algorithm which guarantees the leading order (LO) accuracy of angular correlations between jets is proposed. 
The new algorithm has been constructed with a goal of eliminating the ambiguity in the accuracy of jets, namely it is not necessarily clear whether the kinematics of a jet constructed by clustering partons in final state after a merging procedure has the LO accuracy or not. This ambiguity potentially exists in merging algorithm, because their virtue is smooth combination of LO cross sections and the leading logarithmic (LL) parton shower. 
The new algorithm requires that $n$-jet events are generated exclusively according to the LO $n$-parton production cross section and each of the $n$-jet is close to each of the $n$-parton in terms of the jet measure. As a result, angular correlations between the $n$-jet should follow those between the $n$-parton and thus the LO accuracy of them is robust. Furthermore, as long as the $n$-jet events are exclusively under consideration, only the LO $n$-parton production cross section is needed and hence event generation is efficient. \\

The $t\bar{t}$ production in proton-proton collisions is simulated and then the distributions of the azimuthal angle differences $\Delta \phi=\phi_1^{}-\phi_2^{}$ between the two highest transverse momentum $p_T^{}$ jets with large rapidity separations, namely $|y_1^{}-y_2^{}|>3.5$ and $|y_1^{}-y_2^{}|>2.5$ in addition to $y_1^{} \times y_2^{} < 0$ i.e. the VBF cuts, are studied by using the 2-jet events exclusively. The new algorithm is evaluated by comparing its predictions with the predictions of one of well-established merging algorithms, the CKKW-L algorithm. \\

It has been observed that the new algorithm produces a consistent prediction with the CKKW-L at $n_{\mathrm{max}}^{}=3$, where $n_{\mathrm{max}}^{}$ denotes the maximal number of partons produced by the LO cross section, despite that only the LO $t\bar{t}+2$-parton cross section contributes to the events in the mew algorithm. 
Since the LO accuracy of $\Delta \phi$ is guaranteed in the new algorithm, this observation confirms the previous finding that the LO accuracy of $\Delta \phi$ in the 2-jet events cannot be fully retained due to contribution from the leading logarithmic (LL) parton shower in the CKKW-L at $n_{\mathrm{max}}^{}=2$ and it can be retained by merging the LO $t\bar{t}+3$-parton cross section additionally i.e. in the CKKW-L at $n_{\mathrm{max}}^{}=3$. \\

In the numerical studies of this work, only the $\Delta \phi$ distributions in the $t\bar{t}+2$-jet events are studied and only the CKKW-L algorithm with the PYTHIA8 parton shower is used as one of other merging algorithms to validate the new algorithm. Therefore, the observation in this work is limited to this situation, obviously.  
However, it is always the case that other merging algorithms including the CKKW-L algorithm potentially have the ambiguity in the accuracy of jets, dependently of a process, an observable and a shower model.
In contrast, the new algorithm does not have the ambiguity and always guarantees the LO accuracy of angular correlations between jets, independently of a process, an observable and a shower model. \\

There are basically two parameters which cannot be determined uniquely in the new algorithm, namely $C_{\mathrm{match}}^{}$ and the definition of the parton shower starting scale. 
This fact might be seen as a weak point in the algorithm. However merging algorithms and parton shower programs are just models after all. 
A promising approach will be to tune the algorithm together with a parton shower model by using these parameters so that the $p_T^{}$ and the rapidity distributions of jets in a given process become consistent with the data at first and then make the predictions of observables measuring angular correlations between the jets.

\section*{Acknowledgments}

The work of J.N. is supported by the Institutional Strategy of the University of T\"ubingen (DFG, ZUK 63).



\begin{thebibliography}{99}




\bibitem{Plehn:2001nj}
  T.~Plehn, D.~L.~Rainwater and D.~Zeppenfeld,
  ``Determining the structure of Higgs couplings at the LHC,''
  Phys.\ Rev.\ Lett.\  {\bf 88} (2002) 051801
  [hep-ph/0105325].
  
\bibitem{DelDuca:2001fn}
  V.~Del Duca, W.~Kilgore, C.~Oleari, C.~Schmidt and D.~Zeppenfeld,
  ``Gluon fusion contributions to H + 2 jet production,''
  Nucl.\ Phys.\ B {\bf 616} (2001) 367
  [hep-ph/0108030].

\bibitem{Hankele:2006ma}
  V.~Hankele, G.~Klamke, D.~Zeppenfeld and T.~Figy,
  ``Anomalous Higgs boson couplings in vector boson fusion at the CERN LHC,''
  Phys.\ Rev.\ D {\bf 74} (2006) 095001
  [hep-ph/0609075].

\bibitem{Klamke:2007cu}
  G.~Klamke and D.~Zeppenfeld,
  ``Higgs plus two jet production via gluon fusion as a signal at the CERN LHC,''
  JHEP {\bf 0704} (2007) 052
  [hep-ph/0703202 [HEP-PH]].
  
\bibitem{Hagiwara:2009wt}
  K.~Hagiwara, Q.~Li and K.~Mawatari,
  ``Jet angular correlation in vector-boson fusion processes at hadron colliders,''
  JHEP {\bf 0907} (2009) 101
  [arXiv:0905.4314 [hep-ph]].

\bibitem{Buckley:2010jv} 
  M.~R.~Buckley and M.~J.~Ramsey-Musolf,
  ``Diagnosing Spin at the LHC via Vector Boson Fusion,''
  JHEP {\bf 1109}, 094 (2011)
  [arXiv:1008.5151 [hep-ph]].

\bibitem{Hagiwara:2013jp} 
  K.~Hagiwara and S.~Mukhopadhyay,
  ``Azimuthal correlation among jets produced in association with a bottom or top quark pair at the LHC,''
  JHEP {\bf 1305}, 019 (2013)
  [arXiv:1302.0960 [hep-ph]].






\bibitem{Catani:2001cc} 
  S.~Catani, F.~Krauss, R.~Kuhn and B.~R.~Webber,
  ``QCD matrix elements + parton showers,''
  JHEP {\bf 0111}, 063 (2001)
  [hep-ph/0109231].

\bibitem{Lonnblad:2001iq}
  L.~Lonnblad,
  ``Correcting the color dipole cascade model with fixed order matrix elements,''
  JHEP {\bf 0205} (2002) 046
  [hep-ph/0112284].

\bibitem{Krauss:2002up} 
  F.~Krauss,
  ``Matrix elements and parton showers in hadronic interactions,''
  JHEP {\bf 0208}, 015 (2002)
  [hep-ph/0205283].

\bibitem{Mrenna:2003if}
  S.~Mrenna and P.~Richardson,
  ``Matching matrix elements and parton showers with HERWIG and PYTHIA,''
  JHEP {\bf 0405} (2004) 040
  [hep-ph/0312274].

\bibitem{Lavesson:2005xu}
  N.~Lavesson and L.~Lonnblad,
  ``W+jets matrix elements and the dipole cascade,''
  JHEP {\bf 0507} (2005) 054
  [hep-ph/0503293].

\bibitem{Mangano:2006rw}
  M.~L.~Mangano, M.~Moretti, F.~Piccinini and M.~Treccani,
  ``Matching matrix elements and shower evolution for top-quark production in hadronic collisions,''
  JHEP {\bf 0701} (2007) 013
  [hep-ph/0611129].

\bibitem{Alwall:2007fs}
  J.~Alwall, S.~Hoche, F.~Krauss, N.~Lavesson, L.~Lonnblad, F.~Maltoni, M.~L.~Mangano and M.~Moretti {\it et al.},
  ``Comparative study of various algorithms for the merging of parton showers and matrix elements in hadronic collisions,''
  Eur.\ Phys.\ J.\ C {\bf 53} (2008) 473
  [arXiv:0706.2569 [hep-ph]].

\bibitem{Giele:2007di} 
  W.~T.~Giele, D.~A.~Kosower and P.~Z.~Skands,
  ``A simple shower and matching algorithm,''
  Phys.\ Rev.\ D {\bf 78}, 014026 (2008)
  [arXiv:0707.3652 [hep-ph]].

\bibitem{Alwall:2008qv}
  J.~Alwall, S.~de Visscher and F.~Maltoni,
  ``QCD radiation in the production of heavy colored particles at the LHC,''
  JHEP {\bf 0902} (2009) 017
  [arXiv:0810.5350 [hep-ph]].

\bibitem{Hoeche:2009rj}
  S.~Hoeche, F.~Krauss, S.~Schumann and F.~Siegert,
  ``QCD matrix elements and truncated showers,''
  JHEP {\bf 0905} (2009) 053
  [arXiv:0903.1219 [hep-ph]].

\bibitem{Hamilton:2009ne}
  K.~Hamilton, P.~Richardson and J.~Tully,
  ``A Modified CKKW matrix element merging approach to angular-ordered parton showers,''
  JHEP {\bf 0911} (2009) 038
  [arXiv:0905.3072 [hep-ph]].

\bibitem{Giele:2011cb} 
  W.~T.~Giele, D.~A.~Kosower and P.~Z.~Skands,
  ``Higher-Order Corrections to Timelike Jets,''
  Phys.\ Rev.\ D {\bf 84}, 054003 (2011)
  [arXiv:1102.2126 [hep-ph]].

\bibitem{Lonnblad:2011xx}
  L.~Lonnblad and S.~Prestel,
  ``Matching Tree-Level Matrix Elements with Interleaved Showers,''
  JHEP {\bf 1203} (2012) 019
  [arXiv:1109.4829 [hep-ph]].




\bibitem{Gribov:1972}  
V. N. Gribov and L. N. Lipatov, 
``Deep inelastic e-p scattering in perturbation theory,”
Sov. J. Nucl. Phys. 15 (1972) 438.  

\bibitem{Altarelli:1977} 
G.~Altarelli and G.~Parisi,
``Asymptotic freedom in parton language,” 
Nucl. Phys. B126 (1977) 298–318.

\bibitem{Dokshitzer:1977}  
Y. L. Dokshitzer,
``Calculation of the structure functions for deep inelastic scattering
and e+ e− annihilation by perturbation theory in quantum chromodynamics,”
Sov. Phys. JETP 46 (1977) 641–653.



\bibitem{Hagiwara:2015tva} 
  K.~Hagiwara and J.~Nakamura,
  ``Study on the azimuthal angle correlation between two jets in the top quark pair production,''
  arXiv:1501.00794 [hep-ph].



\bibitem{Sjostrand:2004ef}
  T.~Sjostrand and P.~Z.~Skands,
  ``Transverse-momentum-ordered showers and interleaved multiple interactions,''
  Eur.\ Phys.\ J.\ C {\bf 39} (2005) 129
  [hep-ph/0408302].

\bibitem{Corke:2010yf}
  R.~Corke and T.~Sjostrand,
  ``Interleaved Parton Showers and Tuning Prospects,''
  JHEP {\bf 1103} (2011) 032
  [arXiv:1011.1759 [hep-ph]].

\bibitem{Norrbin:2000uu}
  E.~Norrbin and T.~Sjostrand,
  ``QCD radiation off heavy particles,''
  Nucl.\ Phys.\ B {\bf 603} (2001) 297
  [hep-ph/0010012].


\bibitem{Sjostrand:2007gs}
  T.~Sjostrand, S.~Mrenna and P.~Z.~Skands,
  ``A Brief Introduction to PYTHIA 8.1,''
  Comput.\ Phys.\ Commun.\  {\bf 178} (2008) 852
  [arXiv:0710.3820 [hep-ph]].
  
\bibitem{Sjostrand:2006za}
  T.~Sjostrand, S.~Mrenna and P.~Z.~Skands,
  ``PYTHIA 6.4 Physics and Manual,''
  JHEP {\bf 0605} (2006) 026
  [hep-ph/0603175].


\bibitem{Marchesini:1987cf} 
  G.~Marchesini and B.~R.~Webber,
  ``Monte Carlo Simulation of General Hard Processes with Coherent QCD Radiation,''
  Nucl.\ Phys.\ B {\bf 310}, 461 (1988).


\bibitem{Sjostrand:1985xi}
  T.~Sjostrand,
  ``A Model for Initial State Parton Showers,''
  Phys.\ Lett.\ B {\bf 157} (1985) 321.

\bibitem{Gottschalk:1986bk} 
  T.~D.~Gottschalk,
  ``Backwards Evolved Initial State Parton Showers,''
  Nucl.\ Phys.\ B {\bf 277}, 700 (1986).


\bibitem{Cacciari:2008gp}
  M.~Cacciari, G.~P.~Salam and G.~Soyez,
  ``The Anti-k(t) jet clustering algorithm,''
  JHEP {\bf 0804} (2008) 063
  [arXiv:0802.1189 [hep-ph]].


\bibitem{Alwall:matching}
  J.~Alwall and S.~de Visscher,
  ``Introduction to jet-parton matching in MG/ME,''
  https://cp3.irmp.ucl.ac.be/projects/madgraph/wiki/IntroMatching.

  
\bibitem{Alwall:2014hca} 
  J.~Alwall, R.~Frederix, S.~Frixione, V.~Hirschi, F.~Maltoni, O.~Mattelaer, H.-S.~Shao and T.~Stelzer {\it et al.},
  ``The automated computation of tree-level and next-to-leading order differential cross sections, and their matching to parton shower simulations,''
  JHEP {\bf 1407}, 079 (2014)
  [arXiv:1405.0301 [hep-ph]].

\bibitem{Pumplin:2002vw}
  J.~Pumplin, D.~R.~Stump, J.~Huston, H.~L.~Lai, P.~M.~Nadolsky and W.~K.~Tung,
  ``New generation of parton distributions with uncertainties from global QCD analysis,''
  JHEP {\bf 0207} (2002) 012
  [hep-ph/0201195].
 

\bibitem{Cacciari:2011ma}
  M.~Cacciari, G.~P.~Salam and G.~Soyez,
  ``FastJet User Manual,''
  Eur.\ Phys.\ J.\ C {\bf 72} (2012) 1896
  [arXiv:1111.6097 [hep-ph]].





\end{thebibliography}
\end{document}